# Anisotropic heat conduction in silicon nanowire network revealed by Raman scattering

Mykola Isaiev [1], Oles Didukh [1], Tetyana Nychyporuk [2], Victor Timoshenko [3], and Vladimir Lysenko [2]

[1] *Faculty of Physics, Taras Shevchenko National University of Kyiv, 64/13, Volodymyrska St., Kyiv 01601, Ukraine*

[2] *Université de Lyon; Institut des Nanotechnologies de Lyon, UMR-5270, site INSA de Lyon, Villeurbanne, F-69621, France*

[3] *Lomonosov Moscow State University, Department of Physics, 119991 Moscow, Russia*

**Abstract**

Anisotropic nanomaterials possess interesting thermal transport properties because they allow orientation of heat fluxes along preferential directions due to a high ratio (up to three orders of magnitude) between their in-plane and cross-plane thermal conductivities. Among different techniques allowing thermal conductivity evaluation, micro-Raman scattering is known to be one of the most efficient contactless measurement approaches. In this letter, an experimental approach based on Raman scattering measurements with variable laser spot sizes is reported. Correlation between experimental and calculated thermal resistances of one-dimensional nanocrystalline solids allows simultaneous estimation of their in-plane and cross-plane thermal conductivities. In particular, our measurement approach is illustrated to be applied for anisotropic thermal conductivity evaluation of silicon nanowire arrays.

Thermal transport in nanoscale materials is a very important research topic due to numerous temperature dependent applications of nanomaterials in various fields [1,2]. In particular, precise engineering of heat conduction in various nanomaterials is a crucial issue for fabrication of highly performant nano-devices and nano-systems. For example, significant self-heating of modern devices/systems with high operation speed and power requires efficient ways of heat evacuation from active operational zones in order to avoid any destructive effects induced by overheating. Another example concerns fabrication of highly efficient thermoelectric devices. Indeed, one has to maintain constant temperature difference between "cold" and "hot" poles and an important reduction of the heat fluxes between the poles must be thus ensured for that.

Anisotropic nanomaterials possess especially interesting thermal transport properties because they allow orientation of heat fluxes along preferential directions due to extremely high ratio between their in-plane and cross-plane thermal conductivities. Since this ratio can achieve the level of three orders of magnitudes, such nanomaterials are the excellent candidates for smart distribution of heat fluxes along to preferential directions.

Electrothermal methods, such as 2- and 3-ω techniques, appear to be very efficient for thermal conductivity evaluation of anisotropic materials. For example, this kind of measurements has been recently reported to be applied for revealing of strong thermal anisotropy of nanoporous silicon [3] and 2D materials [4,5]. However, the main disadvantage of the electrothermal methods is related to the use of electric elements which must be in a perfect thermal contact with studied materials or structures. Moreover, huge technological difficulties exist to ensure deposition of these elements on nanostructures with high roughness or aspect ratio. Thus, contactless evaluation of thermal conductivity of nanomaterials is much more desirable.

Since 1999 [6], micro-Raman scattering is known to be an efficient photo-thermal technique for contactless measurement of thermal conductivity of various nanoscale materials [7–9]. This measurement approach uses the same laser beam for local thermal heating of the studied sample as well as for detection of Raman-based light scattering on it. As a result, information on local thermal properties of the studied object (under the area excited by laser) can be extracted.

In this letter, we describe an experimental approach based on Raman scattering measurements with variable laser spot sizes for simultaneous estimation of in-plane and cross-plane thermal conductivities of 1D nanocrystalline solids. In particular, our measurement approach is illustrated to be applied for anisotropic thermal conductivity study of silicon nanowire arrays (SiNWs).

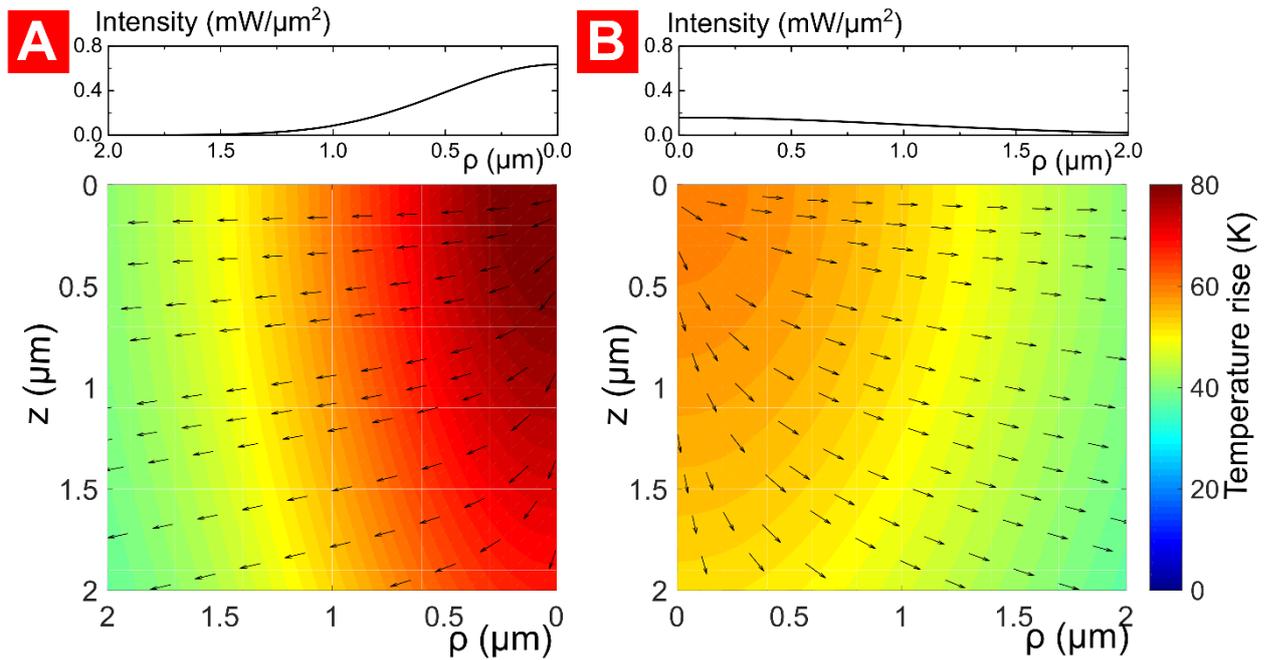

Figure 1. Surface power densities and temperature maps of a thermally isotropic media heated with a laser beam power of 1 mW at z=0μm for two different beam radii: A) 1μm and B) 2μm.

First of all, in order to better illustrate our basic measurement concept, Figure 1 shows temperature maps of a thermally isotropic 20 μm thick sample heated with a laser beam of two different radii (1 μm and 2 μm). The visualized surface power densities and temperature distributions were calculated for identical laser power equal to 1 mW dissipated at z = 0 μm. As one can see, even for the thermally isotropic case, significant difference of heat flux distributions caused by different surface power densities can be clearly stated. In particular, the narrower the laser beam radius is, the more important the vertical component of heat propagation is. Since for an anisotropic material such difference must be even much more pronounced, the Raman scattering measurements performed with different laser beam geometries and correlated with theoretical calculations can be efficiently applied for thermal conductivity evaluation of an anisotropic media.

SiNWs arrays were prepared by metal-assisted chemical etching (MACE) of p-type boron-doped 320 μm thick (100)-oriented bulk crystalline Si (c-Si) substrate. Initial doping level of the substrate was equal to $10^{15}$ cm$^{-3}$. Before the MACE processes, all substrates were rinsed in 5% hydrofluoric acid (HF) aqua solution for 1 min to remove a native silica layer. Then, at the first step, Ag nanoparticles were deposited on c-Si surfaces by immersing them in a solution of 0.02 M silver nitrate ($AgNO_3$) and 5 M HF in the volume ratio 1:1 for 30 s. During the second step, the c-Si wafers covered with Ag nanoparticles were immersed in a mixture of 5M HF and 30% $H_2O_2$ with volume ratio 10:1. The chosen etching times were 35 and 62 minutes and corresponding thicknesses (*l*) of the fabricated SiNWs were, respectively, 20 and 35 microns. Finally, all the samples were rinsed several times in deionized water and dried at room temperature. Additionally, for removing the residual silver nanoparticles from the SiNWs, the formed structures were immersed in 65 % nitric acid ($HNO_3$) for 15 min between the washing processes.

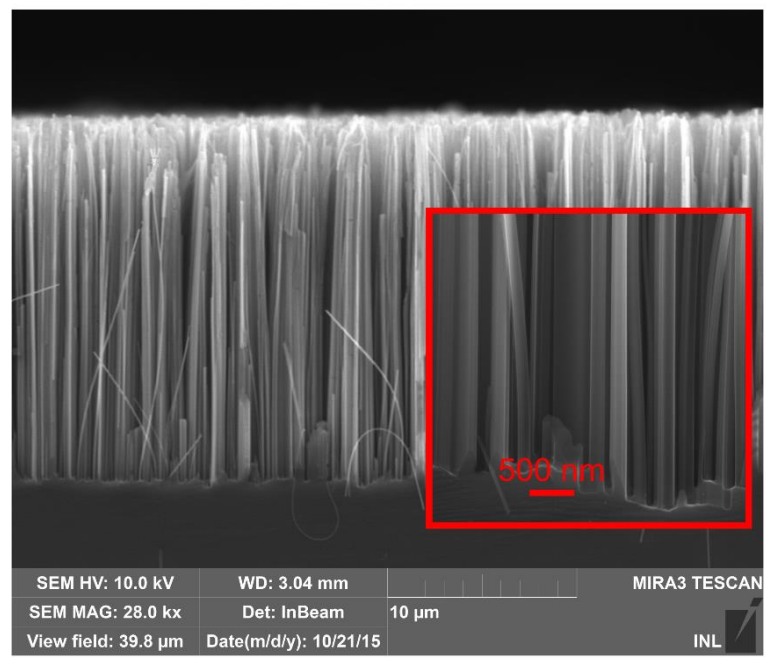

Figure 2. SEM images of the SiNWs obtained by metal-assisted chemical etching during 35 minutes.

Fig. 2 shows typical SEM images of the fabricated SiNWs with diameters being in the range between 80 and 200 nm (with a mean value close to 150 nm). As one can see, the overwhelming majority of them look like quite straight and long vertical rods. However, at the same time, one can see some of the rods to be strongly bent or even broken. As one can easily suppose, such samples with clearly structural anisotropy must be also characterized by corresponding strongly pronounced thermal conduction anisotropy which will be quantitatively described below.

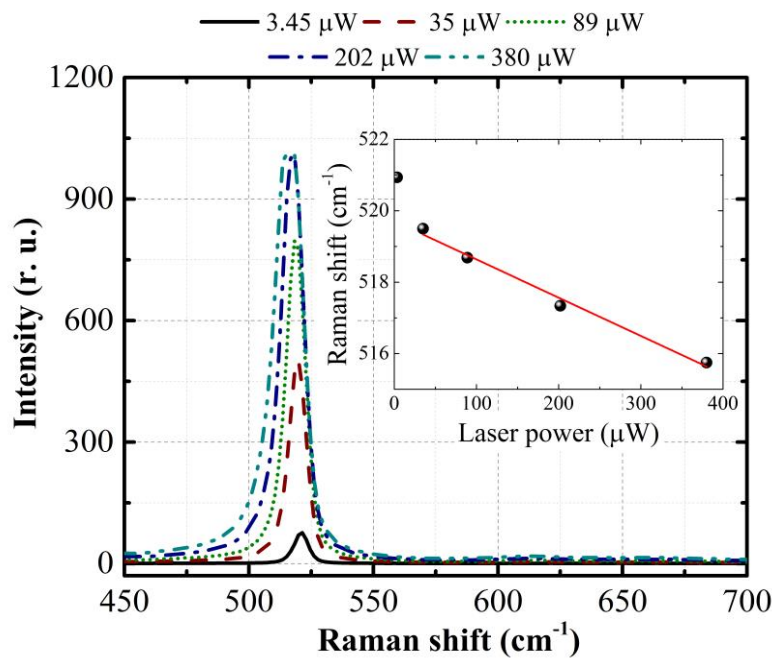

Figure 3. Raman spectra of 35 μm thick SiNWs sample. Laser beam radius was 1 μm. Inset shows dependence of the Raman peak position on laser power.

Dependencies of Raman spectra on laser power in the region from 3 to 400 μW were collected with Renishaw μ-Raman Spectrometer (Fig. 3). Exciting radiation wavelength was 532 nm. Radii of the laser beam (*b*) focused on the sample surface were chosen to be: 1 and 2 μm. The shape and size of the beam profile were monitored by means of a CCD matrix. Then, spectral positions of the peaks and their dependencies on the laser power were determined.

For the temperature rise evaluation, the dependence of the peak position as a function of temperature in a temperature cell was measured. The founded dependence appeared to be linear in temperature range from 290 to 373 K and its slope equal to -0.0232±0.0005 cm$^{-1}$/K is in a good agreement with the data known from literature [7]. Thus, according to the well-known method [6], the slope values of the linear regions of the obtained dependencies (see inset of Fig. 3) were used for evaluation of the thermal resistances ($R = \theta/P$, where $\theta$ is the temperature rise, and $P$ is the laser power) of the studied samples for all the beam radii (the obtained results are summarized in Table 1). The first analysis of the obtained data allows to state an expected decrease of the thermal resistance with increase of the beam radius for a given laser power due to reduction of excitation surface power density.

Table 1. Thermal resistances (K/mW) of SiNWs with two lengths (*l*) evaluated for two laser beam radii (*b*).

|  | $b = 1$ μm | $b = 2$ μm |
|---|---|---|
| $l = 20$ μm | 480 | 170 |
| $l = 35$ μm | 550 | 260 |

For evaluation of thermal conductivity of the studied samples, the following stationary heat conduction equation for the laser induced temperature rise ($\theta$) has been used:

$$\vec{\nabla}(\hat{k}\vec{\nabla}\theta) + s(\vec{r}) = 0, \quad (1)$$

here $\hat{k}$ is the thermal conductivity matrix, $s(\vec{r})$ is the laser induced heating source volume.

For the considered case of silicon nanowires, the matrix of thermal conductivities in a Cartesian coordinate system with the Z axis normal to the sample surface can be written as follows

$$\hat{k} = \begin{pmatrix} k_{in-p} & 0 & 0 \\ 0 & k_{in-p} & 0 \\ 0 & 0 & k_{cr-p} \end{pmatrix}, \quad (2)$$

where $k_{in-p}$ and $k_{cr-p}$ are the in-plane and cross-plane thermal conductivities, respectively.

Absence of a heat outflow from the top sample surface and the thermostating of the bottom one are assumed by the following limit conditions:

$$\begin{cases} \left.(\hat{k}\vec{\nabla}\theta, \vec{n})\right|_{z=0} = 0 \\ \left.\theta\right|_{z=d} = 0 \end{cases}, \quad (3)$$

here $\vec{n}$ is the unit vector normal to the top surface, $d$ is the sample thickness. The equations presented above are written in a coordinate system with the origin located on the sample surface and the Z axis directed in depth of the sample. Additionally, a condition of equality of the heat fluxes and temperature on the interface between SiNWs and bulk silicon substrate (c-Si) was also used:

$$\begin{cases} \left.(\hat{k}\vec{\nabla}\theta, \vec{n})\right|_{z=l-0} = \left.(k_{Si}\vec{\nabla}\theta, \vec{n})\right|_{z=l+0} \\ \left.\theta\right|_{z=l-0} = \left.\theta\right|_{z=l+0} \end{cases}, \quad (4)$$

where $k_{Si}$ is the thermal conductivity of c-Si.

We used a cylindrical coordinate system with the origin at the sample surface and at the epicenter of the laser beam. Expressing gradient and divergence operators in the chosen coordinate system, the equation (1) can be rewritten as follows:

$$\frac{1}{\rho}\frac{\partial}{\partial \rho}\left(\rho k_{in-p}\frac{\partial \theta}{\partial \rho}\right) + \frac{\partial}{\partial z}\left(k_{cr-p}\frac{\partial \theta}{\partial z}\right) + s(\vec{r}) = 0 \quad (5)$$

Assuming that major part of the light intensity is absorbed in the near-surface region of the studied samples, the heat source volume can be expressed in the following form:

$$s(\vec{r}) = \frac{2P_0}{\pi b^2}\alpha e^{-\alpha z} e^{-2\rho^2/b^2} \quad (6)$$

where $P_0$ is the power of laser radiation, $b$ is the beam radius, $\alpha$ is the optical absorption coefficient of the media. For the estimation of optical absorption coefficient of SiNWs array, we use the following equation $\alpha = \alpha_{Si}(1 - P)$, where $P = 0.5$ is the porosity of SiNWs array and $\alpha_{Si} = 7.85 \cdot 10^3$ cm$^{-1}$ is the optical absorption coefficient of crystalline silicon[10].

In addition to the boundary conditions (3) and (4), equation (5) was complimented with the condition of absence of temperature perturbation on a large distance ($R$) from the laser epicenter:

$$\theta(\rho = R, z) = 0, R \gg b \quad (7)$$

This allows us to find a solution of the equation (5) in the following form:

$$\theta(\rho, z) = \sum_{n=1}^{\infty} Z_n(z) J_0\left(\frac{\mu_n^{(0)} \rho}{R}\right), \quad (8)$$

here $Z_n$ is the coefficient of Fourier-Bessel series, $J_0$ is the zero order Bessel function of the first kind and $\mu_n^{(0)}$ is its n-th root.

Substitution of $\theta(\rho, z)$ in the equations (3), (4), and (5) allows us to obtain ordinary differential equation for the coefficients $Z_n$, which can be solved analytically or numerically. Then, one can calculate thermal resistance as $\langle\theta\rangle/P$, where

$$<\theta> = \frac{\int_0^R \theta(0,\rho) \exp\left(-2\rho^2/b^2\right) \rho\, d\rho}{\int_0^R \exp\left(-2\rho^2/b^2\right) \rho\, d\rho} \quad (9)$$

is an averaged temperature [11]. Finally, thermal conductivities of the studied samples can be deduced from correlation between the calculated and experimentally measured thermal resistances. The values of thermal conductivities are obtained from fitting of the experimental results with those obtained from simulation based on equation (9).

Firstly, we have evaluated an average thermal conductivity of the samples assuming isotropic heat conduction in SiNWs media by putting: $k_{\text{SiNWs}} = k_{in-p} = k_{cr-p}$. In Figure 4, the values of averaged thermal conductivity for the 20 μm thick sample is, respectively, presented by big empty (for $b$=1μm) and filled (for $b$=2μm) triangles. As one can state, these values are in good agreement with those already published early [12]. However, difference between the averaged thermal conductivity values evaluated for different beam radii even for the same sample clearly indicate on significant thermal conduction anisotropy of the studied samples.

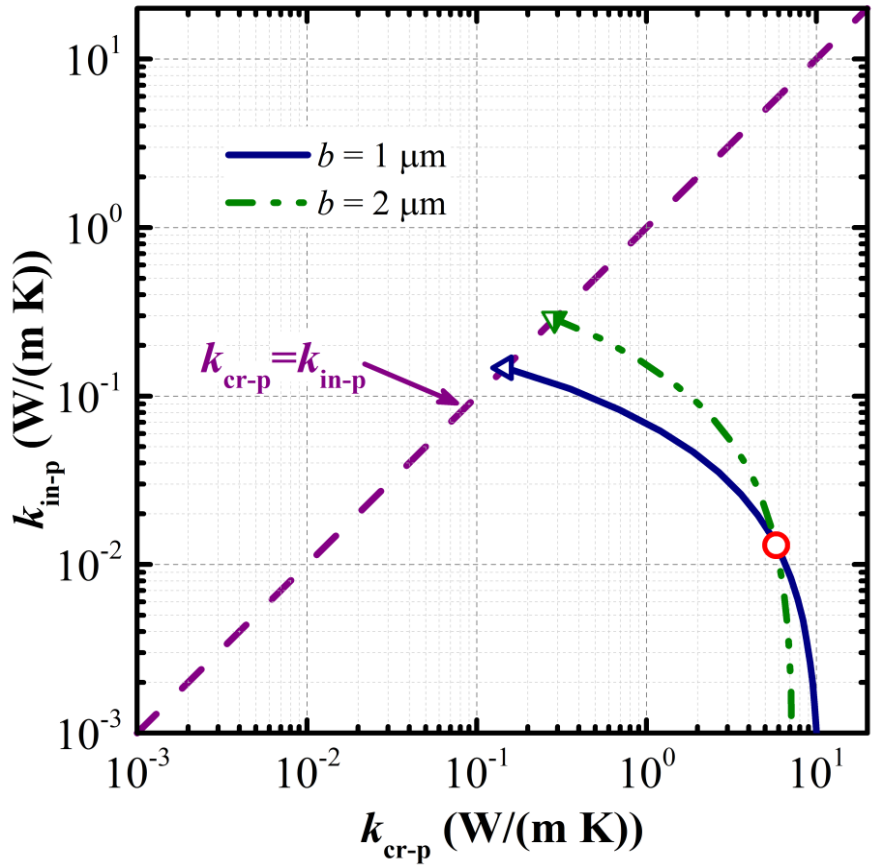

Figure 4. Sets of the pairs $\{k_{cr-p}, k_{in-p}\}$ satisfying experimental thermal resistance values for the 20 μm thick samples. The dashed line corresponds to the assumption of isotopic media.

Since there are two unknown parameters in equation (5) for anisotropic materials, the results of experimental measurements can be fitted with a set of values $\{k_{cr-p}, k_{in-p}\}$. The resulting sets for all the beam radii are presented in Fig. 4 for 20μm thick sample. In order to find the right pair from the whole set, we have proceeded in the following way. Starting from the point $\{k_{\text{SiNWs}}, k_{\text{SiNWs}}\}$ corresponding to the isotropic case (triangles in Fig. 4), we cyclically decrease value of $k_{in-p}$ with some step and simultaneously tuned value of $k_{cr-p}$ to fit experimentally measured thermal resistance with the results of simulations until finding the same pair of values $\{k_{cr-p}, k_{in-p}\}$ for all the beam radii. In Figure 4 this pair corresponds to the intersection point of the represented curves.

Table 2. Thermal conductivity values evaluated from correlation between measured and calculated thermal resistances.

|  | $b = 1$ μm $k_{\text{SiNWs}}$ (W/(m K)) | $b = 2$ μm $k_{\text{SiNWs}}$ (W/(m K)) | $k_{\text{cr-p}}$ (W/(m K)) | $k_{\text{in-p}}$ (W/(m K)) |
|---|---|---|---|---|
| $l = 20$ μm | 0.29 | 0.54 | 5.8 | 0.013 |
| $l = 35$ μm | 0.25 | 0.36 | 4.2 | 0.023 |

The evaluated values of the in-plane and cross-plane thermal conductivities are given in Table 2. As one can see, the reported values of cross-plane thermal conductivity are very close to the values of thermal conductivity evaluated by electrothermal technique for an individual nanowire [13], and have the same order of magnitude as the value evaluated with Time Domain Thermo-Reflectance (TDTR) method for the silicon nanowires with a highly rough surface [14]. The value of the in-plane thermal conductivity is approximately equal to the thermal conductivity of air. Since number of broken nanowire increases with thicnkness of the SiNWs, the cross-plane thermal conductivity of the thinner sample as well as the in-plane thermal conductivity of the thicker one are slightly higher.

In conclusion, an experimental approach based on Raman scattering measurements with variable laser spot sizes has been developed. Correlation between experimental and calculated thermal resistances of one-dimensional nanocrystalline solids allows simultaneous estimation of their in-plane and cross-plane thermal conductivities. Our measurement approach allowed estimation of anisotropic thermal conductivity of silicon nanowire arrays. In particular, the cross-plane thermal conductivity values (4-6 W/(m K)) were found to be more than two orders of magnitude higher than the in-plane ones (0.01-0.03 W/(m K)). The former are in good correlation with the values of thermal conductivity evaluated for a single silicon nanowire [13] and the latter is very close to the thermal conductivity of air separating individual vertical nanowires.